\documentclass[12pt]{article}

\usepackage{amsmath}
\usepackage{times}
\usepackage{natbib}
\font\bf=ptmb8y at 12pt

\makeatletter
\def\cleardoublepage{\clearpage\if@twoside \ifodd\c@page\else
    \hbox{}\thispagestyle{empty}\newpage\if@twocolumn\hbox{}\newpage\fi\fi\fi}
\makeatother

\usepackage{color}
\definecolor{gray}{cmyk}{0,0,0,0.28}

\usepackage[letterpaper]{geometry}
\usepackage{multicol}
\newdimen\stockheight
\newdimen\stockwidth
\stockwidth\paperwidth
\usepackage[center]{crop} 

\usepackage{natbib}
\begin{document}


\title{A General Criterion for Liquefaction in Granular Layers with Heterogeneous Pore Pressure}

\maketitle\pagestyle{empty}


\begin{center}
Liran Goren$^1$, 
Renaud Toussaint$^2$, 
Einat Aharonov$^3$, 
David W. Sparks$^4$, and
Eirik Flekk{\o}y$^5$
\end{center}

\noindent$^1$ETH Zurich, Switzerland; email:  liran.goren@erdw.ethz.ch 

\noindent$^2$Institut de Physique du Globe de Strasbourg (IPGS), UMR 7516, Université de Strasbourg/EOST, CNRS; 5 rue Descartes, F-67084 Strasbourg Cedex, France; email:  renaud.toussaint@unistra.fr

\noindent$^3$Hebrew University, Israel; email:  einatah@cc.huji.ac.il 

\noindent$^4$Texas A\&M University, TX, USA; email:  sparks@geo.tamu.edu 

\noindent$^5$ University of Oslo, Norway; email: e.g.flekkoy@fys.uio.no

\thispagestyle{empty}

\section*{ABSTRACT}
Fluid-saturated granular and porous layers can undergo liquefaction and lose their shear resistance when subjected to shear forcing. In geosystems, such a process can lead to severe natural hazards of soil liquefaction, accelerating slope failure, and large earthquakes. Terzaghi's principle of effective stress predicts that liquefaction occurs when the pore pressure within the layer becomes equal to the applied normal stress on the layer. However, under dynamic loading and when the internal permeability is relatively small the pore pressure is spatially heterogeneous and it is not clear what measurement of pore pressure should be used in Terzaghi's principle. Here, we show theoretically and demonstrate using numerical simulations a general criterion for liquefaction that applies also for the cases in which the pore pressure is spatially heterogeneous. The general criterion demands that the average pore pressure along a continuous surface within the fluid-saturated granular or porous layer is equal to the applied normal stress. 

\section*{INTRODUCTION}
Dynamic loading can induce liquefaction when applied to fluid-saturated granular or porous media. From the macro-mechanical perspective, liquefaction means that the medium transitions from a solid-like material that can sustain some shear stresses without irreversible deformation to a fluid-like material that has no shear resistance. As a consequence, even the smallest shear stresses are relaxed by irreversible flow deformation. In the upper crust of the Earth, liquefaction can occur in response to shear loading, and it is the cause of many natural hazards. Soil liquefaction can take place in response to the passage of earthquake induced seismic waves through a soil column. It causes the soil to lose its shear resistance \citep{Das1993,Kramer1996}, and  as a result, the soil can no longer support the foundations of infrastructures and catastrophic collapse might take place. When slope material collapses under gravity, liquefaction can lead to further acceleration of the sliding mass since the gravitational forces are no longer opposed by the frictional resistance of the sliding body \citep{IversonandLahusen1989}. When fault zones slide during an earthquake, motion is localized along a shear zone that is in many cases filled with granular gouge. Liquefaction of the gouge layer can further enhance the earthquake instability, promoting a longer and faster event \citep{Blanpiedetal1992}. 

Terzaghi's effective stress principle \citep{Terzaghi1943} is commonly invoked to explain the process of liquefaction in such Earth systems. \citet{Terzaghi1943} understood that it is not the applied stress that controls the strength of a solid-fluid system, but instead a quantity termed the effective stress:
\begin{equation}\label{EffectiveStressFull}
\sigma_{ij}' = \sigma_{ij} - \delta_{ij}P,
\end{equation}
where $\sigma_{ij}$ is the applied stress tensor, $P$ is the pressure experienced by the fluid within the pores of a granular or porous material, $\delta_{ij}$ is Kronecker's delta, and  $\sigma_{ij}'$ is the effective stress tensor. According to this principle, the shear resistance, $\tau_c$, of a fluid-filled granular or porous layer is:
\begin{equation}\label{ShearEffectiveStressFull}
\tau_c = \mu(\sigma_n - P),
\end{equation} 
where $\sigma_n$ is the total applied stress normal to the layer, $\mu$ is a friction coefficient, and it is assumed that the material is cohesionless. As a  consequence, when $P=\sigma_n$, $\tau_c = 0$, the layer has no shear resistance, and is will deform irreversibly in response to shear stresses. Such a granular layer is said to be liquefied.   

An important assumption, that is commonly not stated explicitly, is that the pore pressure in the layer is uniform. However, under dynamic conditions, such as simple shear forcing, and when the internal permeability of the layer is relatively small, the pore pressure is not necessarily spatially uniform. In such a case, it is not clear how  Terzaghi's effective stress principle, equation (\ref{EffectiveStressFull}), the shear resistance of the layer, equation (\ref{ShearEffectiveStressFull}), and the criterion for liquefaction should be defined. 

In this work we develop a general theoretical criterion for liquefaction that depends on the pore pressure and applies also when the pore pressure is spatially heterogeneous. We further demonstrate the validity of this criterion using simulations of saturated granular material under forcing of shear at a constant velocity. The general criterion states that a sufficient and a necessary condition for liquefaction of a granular layer that is finite in at least one dimension is that there exist a surface within this layer that separates its boundaries, and along this surface the average pore pressure is equal to the applied normal stress on the layer's boundaries.

\section*{FORCE BALANCE IN FlUID-FILLED GRANULAR MATERIAL}
For most applications, it is sufficient to consider a granular layer that is finite in at least one dimension, i.e. it is bounded in two opposite directions, and the normal stress on its boundaries is uniform. For simplicity, we assume that the two boundaries are perpendicular to the $z$ direction, along which gravity operates. The layer is further assumed to be very long or even periodic (and thus infinite) along the horizontal dimensions, $x$ and $y$. 
We start by writing again the total stress tensor in a fluid-filled granular medium over a representative element volume (REV), with negligible fraction of the surface transmitting the solid stress \citep{Biot1956a}:
\begin{equation}\label{StressTenzor}
\sigma^T = \sigma^s-P\mathbf{I},
\end{equation}
where $\sigma^s$ is the solid stress tensor associated with the forces transmitted between neighboring grains through solid contacts, and $\sigma^T $ is the total stress. Note that $\sigma^s$ and $\sigma^T $ can be identified with the effective stress tensor and the stress tensor from equation (\ref{EffectiveStressFull}), respectively. We use here the notation of total and solid stresses to emphasize the different physical components in the layer. The conservation of momentum statement over a small REV is:
\begin{equation}\label{ForceBalance}
\nabla\cdot\sigma^{T}-\rho_{e}g=\rho_{e}a,
\end{equation}
where $\rho_{e}=(1-\Phi)\rho_{s}+\Phi\rho_{f}$, $\rho_{s}$ is the bulk density of the solid grains, $\rho_{f}$ is the density of the interstitial fluid, and $\Phi$ is the porosity. $g$ is the gravitational acceleration, and $a$ is the total acceleration. Combining equations (\ref{StressTenzor}) and (\ref{ForceBalance}) leads to:
\begin{equation}\label{ForceBalance2}
\nabla\cdot\sigma^{s}-\nabla P-\rho_{e}g=\rho_{e}a.
\end{equation}
We further decompose the pressure, $P$, into a hydrostatic component, $\rho_{f}gz$, and a non-hydrostatic component, $P'$.  Equation (\ref{ForceBalance2}) then becomes:
\begin{equation}\label{ForceBalance3}
\nabla\cdot\sigma_{s}-\nabla P'-(1-\Phi)(\rho_{s}-\rho_{f})g=\rho_{e}a.
\end{equation}

\subsection*{\textit{A general liquefaction criterion}}\label{NewCriterion}
Next, we use the force balance, equation (\ref{ForceBalance3}), together with some assumptions in order to derive a simple and general criterion for liquefaction. The first assumption is that the inertia term, the right hand side of equation (\ref{ForceBalance3}), is relatively small with respect to the terms on the left hand slide. This assumption is mostly valid for slow granular flow, and we revisit it when we analyze the numerical simulations. Equation (\ref{ForceBalance3}) then reduces to a balance of forces:
\begin{equation}\label{ForceBalance4}
\nabla\cdot\sigma_{s}-\nabla P'-(1-\Phi)(\rho_{s}-\rho_{f})g=0.
\end{equation}

To motivate the next stages in the derivation, we note that during liquefaction, shear is not opposed by resisting forces that arise from grain contacts. Therefore, when shear stress or strain is applied to one (or both) of the layer's boundaries, the boundary can slide with no shear resistance only if there exists a continuous surface that separates the two boundaries along which there are no forces arising from grain contacts, i.e. stress chains do not percolate from one boundary to the other. The need for such a continuous surface arises from the non-local nature of the rigidity of granular media, which means that a local region with no solid contact forces can be bypassed by stress chains around it that support the system and maintain resistivity to shear. 

In the following, we will consider a planar surface parallel to the boundaries instead of a general surface. This assumption simplifies the derivation, and it is mostly valid, as we observe in numerical simulations that the separating surfaces responsible for liquefaction tend to be subhorizontal. Again, we discuss the validity of this assumption and its implications in the section dealing with the numerical simulations, below. Since only planes that are perpendicular to the $z$ direction are considered, we focus on the vertical force balance of equation (\ref{ForceBalance4}):
\begin{equation}\label{ForceBalance1D}
\frac{d}{dz}\sigma_{zz}^{s}+\frac{d}{dx}\sigma_{xz}^{s}+\frac{d}{dy}\sigma_{yz}^{s}-\frac{dP'}{dz} - (1-\Phi)(\rho_{s}-\rho_{f})g_z=0,   
\end{equation}
Then, we average equation (\ref{ForceBalance1D}) along a horizontal plane:
\begin{equation}\label{ForceBalanceAverage}
\frac{d}{dz}<\sigma_{zz}^{s}>_h-\frac{d}{dz}<P'>_h-(1-\Phi)(\rho_{s}-\rho_{f})g_z=0, 
\end{equation}
where $<P>_h = (1/A)\int_h P(x)dx$ is the average pore pressure along a plane, $h$, whose area is $A$, and  $<\sigma_{zz}^{s}>_h$ is defined in a similar way. $(1-\Phi)$ in equation (\ref{ForceBalanceAverage}) refers to the porosity above the considered plane. Note, that upon integration, shear stresses on vertical planes are assumed to average out, so the second and third terms in equation (\ref{ForceBalance1D}) become zero. This is the result of the very long or periodic nature of the layer in the $x$ and $y$ directions. Equation (\ref{ForceBalanceAverage}) is then integrated along the 
$z$ direction to give:
\begin{equation}\label{ForceBalanceAverage2}
<\sigma_{zz}^{s}>_h(z)-<P'>_h(z)-(1-\Phi)(\rho_{s}-\rho_{f})g_zz=C,
\end{equation}
Where $C$ is an integration constant.  

The final assumption that  we adopt here is that the gravity term, the third term on the left hand side of equation (\ref{ForceBalanceAverage2}), is relatively small with respect to the total stress. Such an assumption is valid as long as the layer is thin with respect to the overburden that causes the total stress. Using this assumption, $C=\sigma^T_n$, where $\sigma^T_n$ the normal stress on the boundaries of the layer. Equation (\ref{ForceBalanceAverage2}) then reduces to:
\begin{equation}\label{Criterion}
<\sigma_{zz}^{s}>_h(z)-<P>_h(z)=\sigma^T_n, 
\end{equation}
where the prime symbol is dropped from the $P$ as we neglected the hydrostatic contribution to the pore pressure. Note that equation (\ref{Criterion}) is similar to equation (\ref{StressTenzor}), but is valid for quantities averaged over a horizontal plane rather than for a REV. Since solid stresses are always positive, the argument presented before, that liquefaction occurs when there are no solid contact forces across some continuous plane, can be expressed as $<\sigma_{zz}^{s}>_h(z) = 0$ along some plane at depth $z$. Therefore,  equation (\ref{Criterion}) leads to a necessary and sufficient conditions for liquefaction in terms of the pore pressure: 
\begin{equation}\label{Criterion2} 
<P>_h(z)=-\sigma^T_n, 
\end{equation}
which means that along some horizontal plane located at depth $z$ the average pore pressure is equal to the applied normal stress over the boundaries.

\section*{NUMERICAL APPROACH - A COUPLED GRAINS AND FLUID MODEL}
Numerical simulations are used to better illustrate the conditions that are related to liquefaction and in order to validate the theoretical general liquefaction criterion that  we develop herein.  Simulations are performed with a coupled model that combines discrete element method for the granular phase and a continuum fluid solver for the interstitial fluid phase. The details of the coupled approach  are discussed in \citep{Gorenetal2010b,Gorenetal2011}. For clarity, we briefly review it  here. The discrete element component for the solid granular phase implements a 2D granular dynamics (GD) algorithm, in which each individual grain is treated as dynamically tracked particle, subjected to a variety of external forces. Grain interactions, body forces, and the force induced by the interstitial fluid lead to linear and rotational acceleration of the grains:
\begin{equation}\label{NewtonF}
m_i \mathbf{\dot u}_i = m_i \mathbf{g} + \sum_j \mathbf{F}_{ij} - \frac{\nabla P \cdot V_i}{1-\Phi},
\end{equation}
\begin{equation}\label{NewtonT}
I_i \mathbf{\dot w}_i = \sum_j R_{i}\hat{\bf{n}}_{ij}\times\mathbf{F}_{ij},
\end{equation}
where $\mathbf{u}_i$ and $\mathbf{w}_i$ are the translational and rotational velocity vectors of grain $i$ (a superposed dot indicates time derivative). $m_i$ is the grain mass, $\mathbf{g}$ is the gravitational acceleration, $I_i$ is the grain moment of inertia, $R_i$ is the radius of the grain, and $\hat{\bf{n}}_{ij}$ is a unit vector normal to the contact between grains $i$ and $j$. The last term on the right-hand side of equation (\ref{NewtonF}) refers to the force exerted on grain $i$ by the pressure gradient, $\nabla P$, of the fluid surrounding it, normalized by the solid fraction, $(1-\Phi)$, in its vicinity, and by its volume, $V_i$ \citep{Mcnamaraetal2000}. $\mathbf{F}_{ij}$ refers to the inter-granular contact force between grains $i$ and $j$ \citep{Gorenetal2011}. Equations  (\ref{NewtonF}) and  (\ref{NewtonT}), which are solved for each individual grain, give rise to the emergent dynamics of the granular layer.

The formulation for the fluid component is based on mass conservation statements for the solid and fluid phases, Darcy's law, which is a reduced form of the fluid conservation of momentum equation under the assumption of negligible fluid inertia, and a state equation for a compressible fluid. Combined together, a simple three terms equation can be written:
\begin{equation}\label{fluid1}
\dot P - \frac{1}{\beta \Phi \eta}\nabla\cdot[k\nabla P] + 
\frac{1}{\beta \Phi}\nabla \cdot \mathbf{u_s} = 0,  
\end{equation}
where $\beta$ and $\eta$ are the compressibility and viscosity of the pore fluid, respectively. The permeability, $k$, is calculated with a Carman-Kozeny relationship, modified for a 2D volume fraction formulation \citep{Mcnamaraetal2000}, and $\mathbf{u_s}$ is the solid velocity field averaged at a scale over which Darcy's law applied. Equation (\ref{fluid1}) is derived under the assumptions that (1) the compressibility of a grain is negligible relative to the fluid compressibility, (2)  the pore fluid pressure is not too large, i.e. $P\ll 1/\beta$, and (3) the length scale of pore pressure diffusion is always larger than the diameter of a single grain.  The fluid equation (\ref{fluid1}) is solved on a grid that is super-imposed over the granular layer, with grid spacing of about 2 grain diameters. The fluid and solid components interact by mutual interpolation, such that granular packing and grain motion induce interstitial fluid flow and pressure changes, and pore fluid pressure gradients affect the force balance of individual grains. This type of coupling was shown to reproduce pattern formation during sedimentation of grains in liquid \citep{Nieblingetal2010a, Nieblingetal2010b} or gas \citep{Vinninglandetal2007a, Vinninglandetal2007b, Nieblingetal2010b}, and mechanisms of fracture due to fluid injection \citep{Johnsenetal2006,Johnsenetal2007,Nieblingetal2012}.

We use the coupled grains-fluid model to perform simulations of shearing a saturated granular layer at a constant shear velocity. Simulations are performed in a rectangular domain with approximately 1680 (24 $\times$ 70) grains. Grain diameters are drawn randomly from a Gaussian distribution with an average 1 mm, and a standard deviation 1 mm, clipped below 0.8 mm and above 1.2 mm. The simulations are performed at a constant and relatively high total normal stress, $\sigma_n^T$=2.4 MPa, such that the gravity in the thin layer is negligible. Although there is no gravity, we still define a vertical direction, and $\sigma_n^T$ is applied to the two boundaries that are perpendicular to this vertical direction. The layer is periodic in the horizontal direction, and thus analogous to a rotary shear apparatus. 

Each simulation is initiated by compacting a system of loosely packed grains under $\sigma_n^T$, until the porosity equilibrates. We then assume that the pore space is filled with water ($\eta= 10^{-3}$ Pa s, $\beta = 4.5\times 10^{-10}$ Pa$^{-1}$) at zero excess fluid pressure. Variations of pore pressure are measured relative to the initial zero value that corresponds to hydrostatic conditions. For this reason, $\sigma_n^T$ is interpreted as the initial effective stress, under hydrostatic conditions. Finally, a constant shear velocity, $V_{sh} = 0.76$ m/s, is applied to the top wall. During a simulation, $\sigma_n^T$ and $V_{sh}$  are maintained constant, and we measure compaction and dilation in the layer, the shear stress required to shear the top wall at constant velocity, and the evolution of pore pressure. The macro-scale friction coefficient of the system, $\mu_a$, is found by dividing the measured shear stress by the applied $\sigma_n^T$. We report here on two simulations that differ in their drainage boundary conditions: an undrained simulation,  in which there is no fluid flux across the top and bottom boundaries, and a drained simulation, in which the pressure at the top and bottom boundaries is set to zero. 

\section*{LIQUEFACTION IN CONSTANT SHEAR VELOCITY SIMULATIONS}\label{sectionLiquefaction}
\subsection*{\textit{Liquefaction under undrained conditions}}
\begin{figure}[]
\centerline{\includegraphics{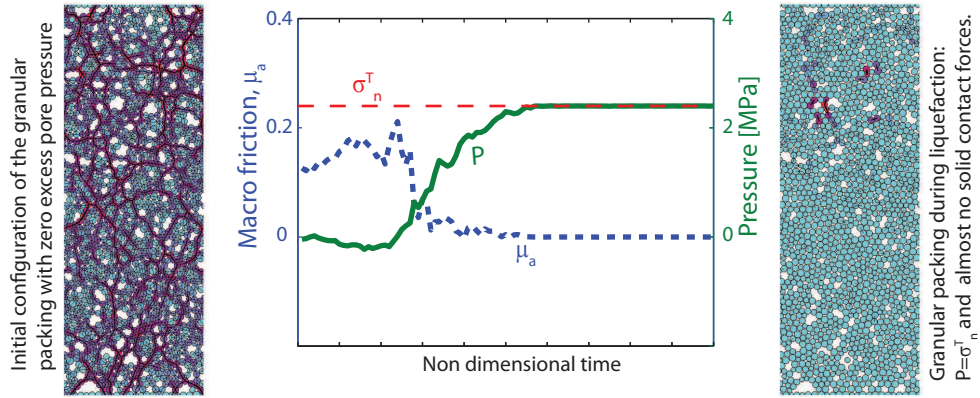}}
\caption{Pore pressure (solid line) and global friction (dotted line) for an undrained saturated layer of grains that compacts during shear. Figures to the left and right show the grain configurations and stresses at the beginning and end of this time period, respectively.}\label{figUndrained}
\end{figure}

First we study an undrained granular layer with an initial high porosity. Such a system undergoes compaction during shear. The internal permeability within the layer is of the order of $k = 10^{-9}$  m$^2$. The time scale related to pore pressure diffusion throughout the layer is defined as $t_d = l \beta \eta \Phi/2k$, where $l\approx$ 70 mm is the height of the granular layer. Due to the high internal permeability, $t_d \approx 3$ $\mu$sec, which is much smaller than the time scale related to shear deformation that is defined as $l/2V_{sh} = 0.05$ sec. Therefore, the pore pressure equilibrates rapidly within the layer and can be considered uniform. Simulation results follow the theory developed in \citep{Gorenetal2010b,Gorenetal2011}, and show that as the pore volume reduces with respect to the initial porosity, the pore pressure rises. Figure \ref{figUndrained} shows that when the pore pressure becomes equal to the applied normal stress $\sigma_n^T$, the solid contact forces disappear, volumetric strain stops growing, the global friction, $\mu_a$, drops to zero, and the fluid saturated granular system has liquefied. Note that here, because of the uniformity of the pore pressure, the classical interpretation of Terzaghi's effective stress principle is valid, and the pore pressure at any point within the system or the average pore pressure within the layer could be used in equations (\ref{EffectiveStressFull}) and (\ref{ShearEffectiveStressFull}), and in Terzaghi's criterion for liquefaction, i.e. $P=\sigma_n^T$. 

\subsection*{\textit{Liquefaction under drained conditions}}
\begin{figure}[h!]
\centerline{\includegraphics{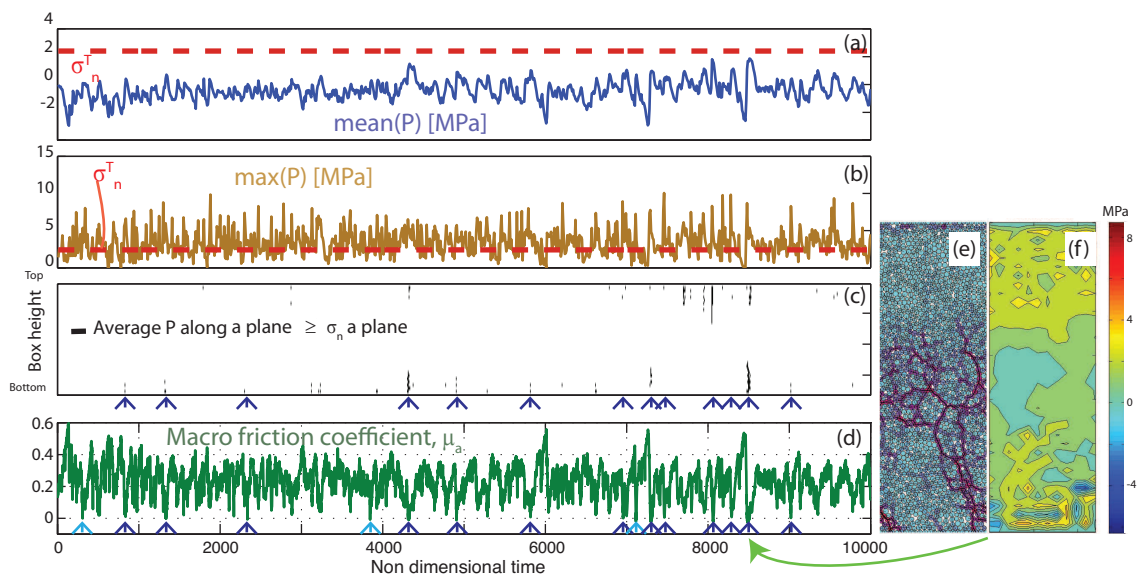}}
\caption{(a)-(d) are time series of pore pressure observables and global friction for a drained grains-fluid simulation. (a) Average pore pressure. (b) Maximum pore pressure. (c) General liquefaction criterion. Locations in which $<P>_h(z) \ge \sigma_n^T$ are marked by plack line. (d) Global friction coefficient. Dark shading arrows show transient liquefaction events that are explained by the general liquefaction criterion. Light shading arrows are transient liquefaction events not explained by the general principle. (e) Granular packing and contact forces, and (f) Pore pressure map during the transient liquefaction event that is marked by the long arrow. Note that the pore pressure is very heterogeneous and that the solid contact forces disappear in localities where the pore pressure is high.}\label{figureProfiles}
\end{figure}

Next, we turn to a drained simulation that starts from a smaller initial porosity and it has a smaller internal permeability of the order of $10^{-14}$ m$^2$. The initial well-compacted configuration leads to an overall dilative trend upon shear, and the smaller internal permeability results in a longer pore pressure diffusion time scale, $t_d\approx 0.3$ sec. As a consequence, the pore pressure cannot equilibrate within the granular layer during the time scale of shear deformation, which is 0.05 sec, and the pore pressure is heterogeneous (see Figure \ref{figureProfiles}f). It might be expected that the dilative trend should prevent generation of high pore pressure, but as shown in \citet{Gorenetal2010b} and \citet{Gorenetal2011}, in drained conditions, $P \propto - \dot\Phi$, so that any small, but rapid, compaction that punctuates the overall dilative trend will lead to positive excess pore pressure. Indeed, despite the dilative trend, the macro friction coefficient, $\mu_a$, reduces to zero or to a small negative value several times during the deformation, in correlation with rapid compaction events. Times with $\mu_a \le 0$ are transient liquefaction events.  Figure \ref{figureProfiles}d, shows the evolution of $\mu_a$ through time. Arrows mark the times when $\mu_a<0.01$ which we regard here as liquefaction. A finite value of 0.01 is used because the record of $\mu_a$ is not continuous during the simulation.

During shear deformation, we cannot define a macro-scale Terzaghi effective stress, and we cannot simply relate a value of $P=\sigma_n^T$ to the liquefaction events because the pore pressure is highly heterogeneous. Instead, we test three different metrics for the spatially variable pore pressure in order to define a criterion for liquefaction. \citet{Gorenetal2011} showed a good correlation between the average pore pressure within the system and the macro-scale friction coefficient, $\mu_a$. However, as can be seen in Figure \ref{figureProfiles}a, pore pressure averaged over teh whole layer, $<P>$, never exceeds $\sigma_n^T$, even during liquefaction events; therefore $<P> = \sigma_n^T$ is not a sufficient criterion for liquefaction. Alternatively, we can consider the maximum value of pore pressure within the system at any given time during shear. Figure \ref{figureProfiles}b shows that the maximum value of pore pressure exceeds $\sigma_n^T$ for almost 70\% of the simulation duration, and therefore a criterion of the form $max(P) =\sigma_n^T$ is not a necessary condition for liquefaction.

As a third metric, we calculate $<P>_h$ along all horizontal planes within the system through time. Figure \ref{figureProfiles}c shows the locations along the layer's  height where $<P>_h \ge \sigma_n^T$ during the simulation.  Arrows mark the times when such a condition is true and the friction curve drops to a value smaller than 0.01. The correlation between the occurrences of $<P>_h \ge \sigma_n^T$  and $\mu_a < 0.01$ is good, and supports the validity of our general liquefaction criterion, equation (\ref{Criterion2}). For the times in which the friction is low but $<P>_h(z) < \sigma_n$ for all $z$ (light-color arrows in Figure \ref{figureProfiles}d), we speculate that a non-horizontal surface, $s$, for which we do not account here satisfies $<P>_s \ge \sigma_n$. For the times in which $<P>_h \ge \sigma_n$ but the friction do not drop to low enough value, we offer that inertia is non-negligible, and therefore the general criterion is not entirely valid. Note that also when $<P>_h > \sigma_n^T$, the  inertia is not negligible, but this does not affect the predictive nature of a general liquefaction criterion of the form $<P>_h \ge \sigma_n^T$.

\section*{CONCLUSION}
Terzaghi's principle of effective stress can be cast as a criterion for liquefaction along a fluid-filled granular layer when, $P=\sigma_n^T$. However, this criterion is valid as long as the pore pressure is spatially uniform. Under dynamic loading and when the permeability is relatively small, the pore pressure is spatially heterogeneous and therefore the classical criterion is not useful because it doesn't define where the pore pressure should be measured. 

We develop here a general criterion that is suitable also for systems with heterogeneous distribution of pore pressure. The criterion states that a necessary and sufficient condition for liquefaction is the existence of a continuous surface along which the average pore pressure is equal or greater than the applied normal stress on the boundaries of the layer. According to this criterion, the value of pore pressure at local points or the average value of pore pressure as measured in experimental or natural settings are not sufficient for relating the pore pressure to loss of shear resistance. Instead, a full pore pressure map is required and an analysis of the general criterion is needed.



\end{document}